\documentclass[journal
                ]{IEEEtran}
\usepackage{graphicx}
\usepackage{longtable}
\usepackage{amsmath}
\def\l#1{\label{eq:#1}}
\def\r#1{(\ref{eq:#1})}

\def\e{\begin{equation}}
\def\f{\end{equation}}

\renewcommand{\baselinestretch}{1.64}

\title{Improving antenna near-field pattern by\\ use of artificial impedance screens}
\author{Stanislav Maslovski,~\IEEEmembership{}
Pekka Ikonen,~\IEEEmembership{Student Member,~IEEE,}
Constantin Simovski,~\IEEEmembership{}\\
Mikko K\"{a}rkk\"{a}inen,~\IEEEmembership{}
Sergei Tretyakov,~\IEEEmembership{Senior Member,~IEEE,} and
Vasil Denchev~\IEEEmembership{}%
\thanks{Stanislav Maslovski, Pekka Ikonen, Mikko K\"{a}rkk\"{a}inen, and Sergei Tretyakov are with Radio Laboratory$\,/\,$SMARAD,
Helsinki University of Technology,
P.O. Box 3000, FIN-02015 HUT, Finland. Contact e-mail: {\tt stanislav.maslovski@hut.fi}}%
\thanks{Constantin Simovski is with Physics Dept. of the St. Petersburg State University of Information Technologies, Mechanics and Optics, 14 Sablinskaya str., 197101, Saint-Petersburg, Russia. Contact e-mail:
{\tt simovsky@phd.ifmo.ru}}}

\begin{document}

\maketitle

\begin{abstract}
An antenna prototype utilizing artificial impedance surfaces
to control the near field distribution is described. The antenna
is a folded dipole placed above a finite-size artificial impedance
surface. We have found that the field screening is most effective
if the surface is a metal conductor. However, to achieve a reasonable value of the radiation resistance
the dipole should be located far off the screen. If the surface is a magnetic wall, the antenna
design is more compact, but the field behind the screen is large.
Here we realize a compromise solution using an inductive surface of a moderate surface impedance,
which allows realization of an effective near-field screen with still a reasonably
low-profile design.
\end{abstract}


\section{Introduction}
In paper \cite{1} an idea of decreasing the field level on one
side of low-profile antennas was proposed and developed. Using
such antennas, efficiency can be maintained at a good level even
if absorbing bodies are near the antenna but behind the screening
substrate, because the power absorbed in the near vicinity of the
antenna is reduced. The idea is based on the usage of artificial
impedance surfaces with moderate inductive impedances to form the
desired near-field pattern of a horizontal antenna positioned
parallel to the impedance surface. The known solutions for
artificial antenna substrates (e.g., \cite{2}) utilize impedance
surfaces in the resonant regime (as artificial magnetic walls) in
order to reduce the antenna thickness. In papers \cite{2}--\cite{5}
implied applications of artificial impedance surfaces are within
the resonant band. However, it has been shown in \cite{1,6} that
the antenna solutions with a high inductive impedance of the
ground surface do not correspond to a good screening effect in the
near-zone field of horizontal antennas. The present approach
allows an effective reduction of the near field behind the
antenna, whereas the required radiation resistance can be achieved
with a small increase of the structure thickness (the primary
radiator antenna is positioned at a certain height over the
impedance surface). Impedance screens of two kinds have been
studied. The first one is so-called \textit{mushroom} structure
and the second one is a thin layer named \textit{Jerusalem
crosses} structure. These impedance screens should work well when
the surface is of moderate inductive impedance and when the
antenna excites mostly TE modes (respectively to the screen
normal). In this work we validate experimentally our theoretical
expectations designing optimized radiating elements for the use
with two different impedance surfaces and measuring the near field
distributions around the antenna. We do not study the
far-field pattern since it is not a practically important
characteristic for handsets operating in the frequency range 1--2
GHz. Really, this pattern is weakly directive (because the
antenna of a mobile phone is electrically small), and it is
perturbed depending on the device position with respect to the
user. However, the non-perturbed field distribution measured in
the vicinity of a portable terminal clearly indicates the
possible influence of the user body or other nearly located
objects to the antenna. If the field
behind the terminal (where the user's body is normally) is not significant,
the user's body will not significantly reduce the antenna efficiency.
Therefore, we make measurements at the distances of a few
centimeters to test the near-field screening efficiency of the artificial
impedance surface.

Notice, that in the known literature the problem of the improving the
antenna efficiency and SAR
reduction is often considered in terms of far-zone measurements and
related with such a parameter of the antenna as the front-to-back
ratio in the far zone. Following this point of view, if one wants to reduce the
SAR one should form the far-zone antenna pattern with a small
backward radiation \cite{Pisa,7}. This is a potentially misleading
approach.
There is no proportionality between the far-zone and the near-zone
field patterns, and a high front-to-back ratio for the far-zone fields
can correspond to a small front-to-back ratio in the
near zone. Only the measurements of the near-field spatial
distribution (and not the field angular dependence since the
angular near-field pattern strongly depends on the distance from the
antenna) can definitely indicate the degree of field interactions
between the antenna and various objects in its near field. Our measurements are,
therefore, not conventional and they became possible due to the
use of special equipment dedicated for near-field testing.

In the proposed design, the artificial impedance surface (such
as the mushroom structure) is operating at frequencies below
the surface impedance resonance, where the effective surface
impedance is inductive and not high in the absolute value.
In this frequency range the surface supports
TM-polarized surface waves, whose excitation is not desirable.
This means that we should use a primary radiator that excites
(at least primarily) TE-polarized waves, that is, there should
be mainly electric currents parallel to the surface. A natural
choice is a simple dipole antenna positioned parallel to the
surface. Another choice criterion is the compact design. When
a dipole antenna is brought close to a surface with a small inductive
impedance, its radiation resistance becomes rather low due to
cancelation of radiation from the currents induced on the surface.
For this reason, a better choice is a folded dipole because
of its high radiation resistance.

The use of a folded dipole requires a balun to connect the dipole
to the unbalanced coaxial cable. At the first stages of our experiments
we wanted to avoid complications related to any additional
elements in the antenna system, and decided to use in experiments
only a half of the dipole antenna (over a half of the artificial
impedance surface) positioned orthogonal to a large metal ground
plane. The mirror image of the half of the antenna simulates
the other half, so that the whole system radiates as the complete
antenna in the half space where the actual radiator is located.
No balun is needed if the feeding cable is behind the ground
plane because no currents are induced on the cable.

After successful tests with this installation, we designed a
planar balun, manufactured and tested a complete antenna sample.
Let us emphasize that our paper is not a description of a new practical antenna
with good front-to-back ratio in the far field.
The goal of this paper is to demonstrate the
feasibility of handset antennas for the frequency range 1--2 GHz with
reduced interaction with objects in the antenna near field region.

\section{Modeling measurements and simulations with one half of the antenna structure}
\label{half_dip}
\subsection{Experimental set-up}

\begin{figure}[htbp]
\begin{center}
\vspace{3cm}
\includegraphics[width=8cm]{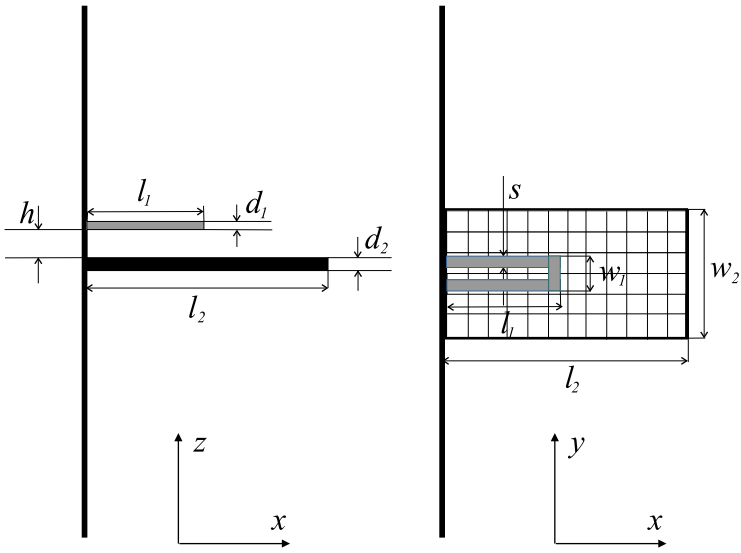}
\vspace{2.3cm}
\end{center}
\caption{Experimental modeling set-up: side view (left) and top view (right).}
\label{modeling_setup}
\end{figure}

The experimental set-up modeling the complete antenna is shown in Figure~\ref{modeling_setup}. It consists of a $220\times 230$ mm$^2$ rectangular
metal screen with a 50 Ohm 3.5 mm coaxial connector in its center
and a half of the antenna structure attached to the connector
(and also to the screen, as explained below).

\begin{figure}[htbp]
\begin{center}
\includegraphics[width=8cm]{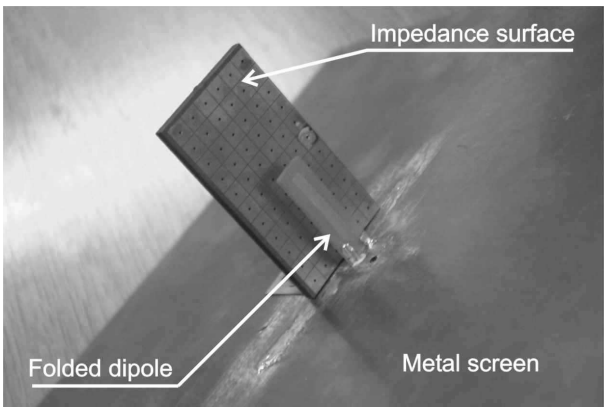}

\end{center}
\caption{Photo of the modeling set-up.}
\label{model_setup_photo}
\end{figure}

Neglecting the influence of the finite sizes of the screen and
the effect of the finite screen conductivity, a half of the real
antenna and its mirror image can be considered as forming the
entire antenna. The source voltage (the source is represented
by the coaxial connector in our case) is then applied between
the screen and one end of one half of the folded dipole. The
second end of the dipole is connected to the ground (see Figure~\ref{model_setup_photo}). The impedance surface should also be well electrically connected
to the ground.

Cabling effects are eliminated here due to the simple fact that
the cable is now behind the screen. It prevents currents from
flowing on the outer surface of the coax. This means that there
is no need for a special balancing device which can possibly
influence the input impedance measurements, etc. The input impedance
of the complete antenna is two times larger than that measured
for a half of the antenna in this experimental set-up. The field
distributions measured in this modeling set-up should correspond
to the same distributions for the complete antenna in free space.
The position of the impedance surface is adjustable mechanically,
and the separation between the dipole and the surface can be
changed. Three types of surfaces have been used in the experiments:
a mushroom structure, a metal screen with slits shaped as Jerusalem
crosses, and a simple metal screen, all of the same surface dimensions.

Antenna dimensions were chosen for operation around 1.8~GHz.
To take into account the influence of the impedance screen
on the effective resonant length of the antenna, measurements
with a test microstrip transmission line formed above the impedance screen have been made. It was found that
in the presence of the impedance surface the resonant length
is slightly shorter than the same for the antenna in free space.
It has been set approximately to 30 mm (it is the dipole half-length
corresponding to the quarter wavelength distance). Folded dipole
samples have been prepared from a metallized dielectric material
(sheets of \textit{FR-4}). Geometrical parameters for all studied
cases (sizes are in mm, for the definition see Figure~\ref{modeling_setup}) are
given in Table~\ref{tab1}.
The last column gives the total thickness of
the structures: $H = d_1+d_2+h$. Distance \textit{h} between the folded dipole and
the impedance surface determines the input impedance of the antenna.
The following experimental procedure to find an optimal distance
was used. The working frequency was firstly chosen (close to 1.8~GHz).
Then, the distance between the folded dipole and the impedance
surface was changed until the imaginary part of the antenna input
impedance became zero at the working frequency. If the real
part of the impedance is close to 25 Ohm (note that in the modeling
set-up we measure one half of the actual impedance of the complete
antenna) for the same frequency, then the antenna is matched
and ready for measurements. If this condition was not reachable,
then a frequency close to the original one was chosen and the procedure was
repeated.

\begin{table}[h]
\begin{center}
\caption{Dimensions of the antenna elements (in mm).}
\label{tab1}
\begin{tabular}{|l|l|l|l|l|l|l|l|l|l|}
\hline
                  & $l_1$ & $l_2$ & $d_1$ & $d_2$ & $h$ & $w_1$ & $w_2$ & $s$ & $H$\\
\hline
Mushrooms         & 33    & 62    & 1.6   & 3.3   & 4.6 & 9.0   & 32    & 3.0 & 9.5\\
\hline
Metal             & 33    & 62    & 1.5   & 0.45  & 7.4 & 10    & 30    & 3.2 & 9.35\\
\hline
Jerusalem         & & & & & & & & &\\
crosses& 33    & 60    & 1.5   & 0.80  & 10  & 10    & 33    & 3.2 & 12.3\\
\hline
\end{tabular}
\end{center}
\end{table}

\subsection{Impedance surfaces}

Two designs of artificial impedance surfaces have been experimentally
studied and compared with the case of a metal plate of the
same size. The sizes of the impedance surfaces in both variants
are indicated in Table~\ref{tab1}.
Design 1 (mushroom structure) is an
array of square metal patches on the upper surface of a metal-backed
dielectric layer. The central points of every patch are connected to the
ground plane by vertical vias. The theory of mushroom structures
is well-known \cite{3,8}. The parameters of mushrooms and the dielectric
layer (\textit{Taconic TLY-5}) are shown in Figure~\ref{impedance_surf_right}. The
surface impedance of this mushroom structure at $f=1.8$ GHz
was theoretically estimated as $Z_s\approx j50$ Ohm. We used the analytical
theory from \cite{8} which gives following approximate relations for a
simple mushroom structure:
$$
Z_s\approx j{\omega L\over 1-\omega^2LC},
$$
\begin{equation}
L = \mu_0 d_2, \quad C = \frac{D\varepsilon_0(\varepsilon_r + 1)}{\pi}\log\frac{2D}{\pi\delta},
\l{ms}\end{equation}
where $d_2$ is the thickness of the mushroom structure,
$D$ is the patch array period and $\delta$ is the gap between patches.

\begin{figure}[htbp]
\begin{center}
\includegraphics[width=6cm]{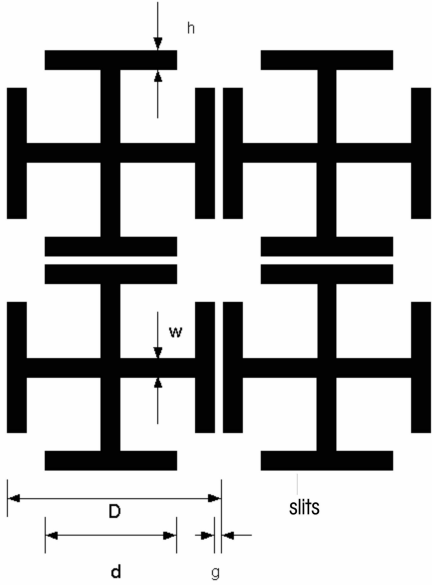}
\end{center}
\caption{A grid of slots shaped as Jerusalem crosses (slots are in black).}
\label{impedance_surf_left}
\end{figure}

\begin{figure}[htbp]
\begin{center}
\vspace{3mm}
\includegraphics[width=3cm]{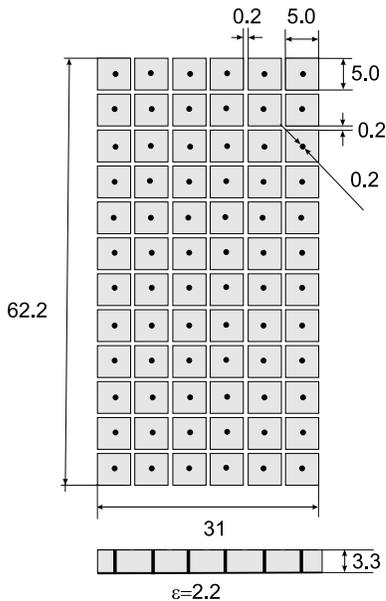}
\end{center}
\caption{Mushroom surface and dimensions of its elements (in mm).}
\label{impedance_surf_right}
\end{figure}

Notice, that a moderate inductive impedance is needed for
the expected screening effect \cite{1,6}, and the obtained result for
$Z_s$ fits to this condition. Design 2 (a screen with
complex-shaped slits) is formed by a grid of slits made in a
metal covering of a thin dielectric substrate (of the relative permittivity $\varepsilon_r=4.5$).
The thickness of the substrate is equal to 0.8 mm. The notations
for all the dimensions of the slits are given in Figure~\ref{impedance_surf_left}.
On this figure the slits in the metal screen are shown in black. The following sizes were chosen for the experimental sample: $g=w=0.2$ mm, $d=2$ mm, $D=4.2$ mm, $h=0.8$
mm. The theory of such surfaces is known for both cases
when the substrate is metal-backed and when it is free \cite{4,5}. In our case it is free, and the surface impedance
at low frequencies is complex with inductive imaginary part.
It can be found as a parallel connection of the grid impedance of the slotted screen $Z_g$ and the input impedance of the dielectric layer of thickness $d_2$:
$$
Z_s={Z_dZ_g\over Z_d+Z_g}.
$$
The grid impedance (relating the tangential electric field in the screen plane
and the surface current induced on it) can be approximately presented as
\begin{equation}
Z_g\approx j{\omega L_g\over 1-\omega^2L_gC_g},
\l{new}\end{equation}
and the dielectric layer in free space has the following surface impedance:
\begin{equation}
Z_d=\eta{1+{j\over \sqrt{\epsilon_r}}\tan kd_2\over 1+{j\sqrt{\epsilon_r}}\tan kd_2}.
\l{new1}\end{equation}
In \r{new} the effective inductance and capacitance determining the {\it grid} impedance of the uniplanar screen can be expressed as
\begin{equation}
L_g = \frac{\mu_0d}{\pi}\log\frac{2d}{\pi g}, \quad C_g = \frac{D\varepsilon_0(\varepsilon_r + 1)}{\pi}\log\frac{2D}{\pi w}.
\end{equation}
The notations are clear from Fig. \ref{impedance_surf_left}.
In \r{new1} $k=\omega\sqrt{\varepsilon_0\mu_0\varepsilon_r}$ is the wave number of the dielectric medium. The real part of $Z_s$ corresponds to the penetration of radiation through the impedance surface. For this design the analytically
estimated surface impedance was $Z_s\approx 0.80+j17$ Ohm at $1.8$ GHz.
This is also a moderate surface impedance.

To understand the influence of such impedance surfaces  to
the input parameters of the folded dipole we have calculated the
wave impedance and the propagation constant of the infinite metal
strip of width $3-4$ mm raised at $5-10$ mm over the impedance
plane. This was done with the exact image method
\cite{Lin}. The results have shown no dramatic
influence, and this has been confirmed by measurements.

\subsection{Measurements}

Measurements with the modeling set-up have been done for a folded
dipole above three types of impedance surfaces: metal surface,
mushroom structure, and Jerusalem crosses structure. Near fields
have been measured by a special set-up (proprietary design of SPEAG and HUT Radio Laboratory) developed for SAR measurements. The set-up has two probes,
electric and magnetic, which allow measuring the \textit{amplitudes} of
three components of the field vectors. Unfortunately, the dipoles
in the electric probe are oriented differently relatively to
the usual laboratory coordinate system: They are \textit{not} along
the two horizontal directions and the vertical one. Because such
components cannot be transformed to the usual ones without knowledge
about the phases of the fields, the total absolute values of
the field vectors have been measured.

At first, the near field distribution over the radiating
folded dipole was measured, to check that the modeling antenna was operating in the desired regime
and the field distribution corresponded to the expected pattern.
The probe moved in a plane 7 mm above the half-dipole (here \textit{above} corresponds to the orientation shown
on Figure~\ref{modeling_setup}, left). The measurement points covered an area equal to the area of the underlying mushroom structure.
The measurements showed that the electric near field had
a maximum at the end of the half-dipole, and the magnetic field had its maximum close to the feeding point, as expected.

\begin{figure}[htbp]
\begin{center}
\includegraphics[width=8cm]{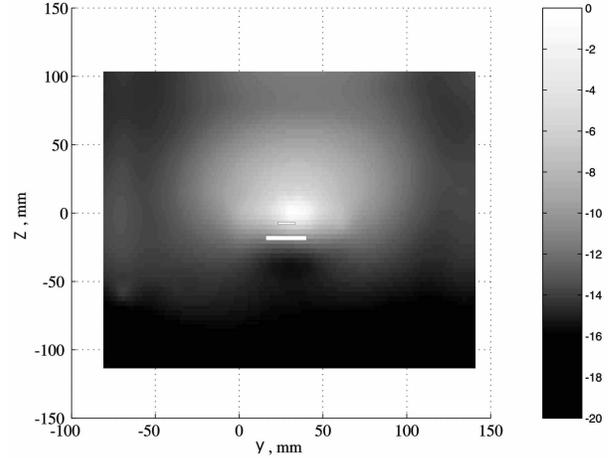}

\end{center}
\caption{Near-field spatial distribution in the vertical plane across the horizontally positioned folded half-dipole located over the mushroom impedance screen (electric field vector relative magnitude, dB).
The antenna and the ground plane location are shown.
The picture top corresponds to the region in front of the antenna, the bottom represents the screened area. The antenna structure is located in the region $0<y<70$ mm,\break $-40\ \mbox{mm}<z<0$.}
\label{fig6}
\end{figure}

\begin{figure}[htbp]
\begin{center}
\includegraphics[width=8cm]{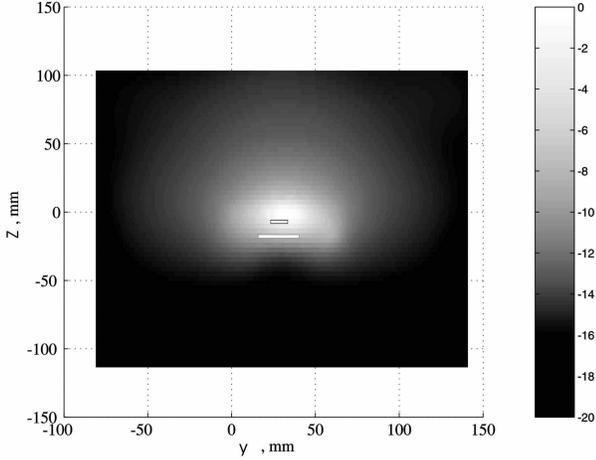}

\end{center}
\caption{The same as in Figure~\ref{fig6} for the magnetic field relative magnitude, dB.}
\label{fig7}
\end{figure}

Next, we measured the distribution of the near field in
vertical planes which included the antenna cross section. In the
following series of measurements the probe was moved around the
antenna along a planar spiral path in the $yOz$) plane (the axes
are accordingly to Figure~\ref{modeling_setup}). The path covers
an area of $220\times 230$ mm$^2$. The measurements have been
performed at the antenna resonant frequency. The electric near
field pattern of the folded dipole placed over the mushroom
surface is depicted on Figure~\ref{fig6}. Figure~\ref{fig7} shows
the magnetic near-field distribution over the same plane.
The field values are given in dB relatively to the maximum level.
The coordinate axes are parallel to the edges of the ground plate.
In these figures the plane over which the field distributions
are measured is located in the middle of the half-dipole. The
bottom side of the picture corresponds to the area behind the
impedance surface where the screening effect is significant. The
region $0<y<70$ mm, $-40\ \mbox{mm}<z<0$ should be excluded from
the plots. This region was occupied by the antenna structure and
it was impossible to move probes inside this area. As a
result, we have measured the near field distribution
inside a rather large spatial box containing the antenna except a small box
having the sizes of the antenna itself.

We define the local screening effect (LSE) as the ratio (in dB) between the field amplitudes at two
points located in front ($z>5$ mm) and behind ($z<-45$ mm) the antenna structure, equidistantly
from the antenna center. The averaged screening effect (ASE)
is the averaged value of LSE over all these points covered
by our measurements. For the mushroom structure ASE is approximately
equal to 15 dB for electric fields and 20 dB for magnetic fields.

Next, we have studied the near-field patterns of a half of the
folded dipole when the mushrooms are replaced by a metal surface and by the Jerusalem-crosses
surface. The screening effect is maximal for the metal surface: ASE is close to 20 dB for both
electric and magnetic fields. But for the input impedance of the half-dipole to be close to 25 Ohm,
the metal screen should be positioned at a larger distance from the
source than the mushroom layer (parameter $h$ in Table~\ref{tab1}).

To compare the radiation properties of the studied antennas measurements of antenna impedances
and antenna efficiencies have been done. The input impedance measurements have been performed with the
HP 8753D network analyzer. To measure the efficiency, the antenna
samples were covered by a conducting semi-sphere (Wheeler's cap),
and the real part of the input impedance was measured. Then,
the efficiency value was calculated as

\begin{equation}
\eta =(1-R_{0} /R)\cdot 100\%
\end{equation}
where \textit{R}$_{0}$ is the real part of the input impedance of the
covered antenna at resonance, and \textit{R} is the real part of the
antenna impedance without the covering, at the resonance. The results of measurements are given in Table~\ref{tab2}.
The best efficiency was achieved with the Jerusalem crosses structure.
But in the same time this structure was the thickest one (see
Table~\ref{tab1}).

\begin{table}
\begin{center}
\caption{Wheeler-cap antenna efficiency measurement results for the antennas under test.}
\label{tab2}
\begin{tabular}{|l|l|l|l|}
\hline
               & Mushrooms & Metal & Jerusalem crosses\\
\hline
$R$, Ohm       & 22        & 25    & 25\\
\hline
$R_0$, Ohm     & 6         & 6     & 3\\
\hline
Efficiency, \% & 73        & 76    & 88\\
\hline
\end{tabular}
\end{center}

\end{table}

\subsection{Some numerical results by the FDTD method}

\begin{figure}[h]
\vskip5mm
\begin{center}
\includegraphics[width=7cm]{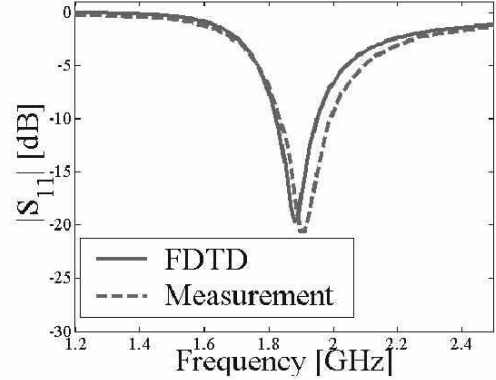}
\end{center}
\caption{The magnitude of \textit{S}$_{11}$ parameter of the
folded dipole over the impedance surface. The measured and the
simulated results agree relatively well.}
\label{Mikko}
\end{figure}

A three-dimensional computer code for the numerical calculation
of the input antenna parameters has been developed. The FDTD method
was used to solve the Maxwell equations in the time domain, the
simulated data were converted into the frequency domain by the
Fourier transform. The mushroom surface impedance was modeled as the input
impedance of a parallel circuit with the parameters
given by \r{ms}. For details of the simulation method,
see \cite{6}. In this example the distance of the dielectric support
of the radiator from the impedance screen is 7.7 mm. The numerical and
measured results for the magnitude of \textit{S}$_{11}$ parameter are
presented in Figure~\ref{Mikko}.
The smallest reflection occurs at about
1.9 GHz, in agreement with the measured results.
So, the antenna is tuned as we need and has a rather good bandwidth for this frequency range ($8\%$ on the level $-10$ dB).

\section{Full-sized folded dipole and a symmetrized prototype antenna}
\subsection{The prototype antenna}
\begin{figure}[htbp]
\begin{center}
\includegraphics[width=7cm]{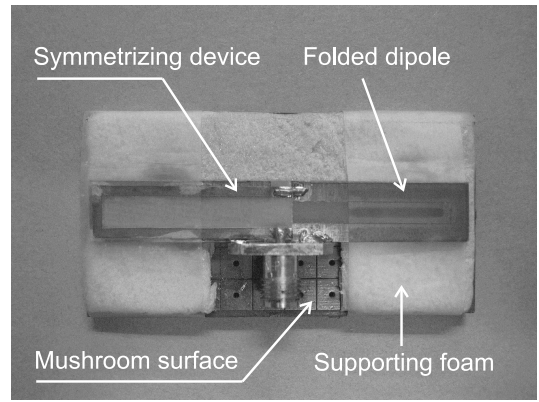}
\end{center}
\caption{The prototype folded-dipole antenna with an integrated symmetrizing device. Behind the antenna a foam layer and the mushroom surface are seen.}
\label{foldeddip_proto}
\end{figure}

Experiments with the modeling set-up (Section \ref{half_dip}) have shown
that there is a possibility to obtain a 50 Ohm (full-sized) matched antenna
using the folded dipole placed on top of an impedance surface.
The needed distance of the antenna from the impedance surface
was found experimentally. When a real full-size antenna is feeded
by a coaxial cable, a symmetrizing device is needed.
Such a device has been designed and manufactured using the
planar technique. The photo is shown in Figure~\ref{foldeddip_proto}.

\begin{figure}[htbp]
\begin{center}
\includegraphics[width=5cm]{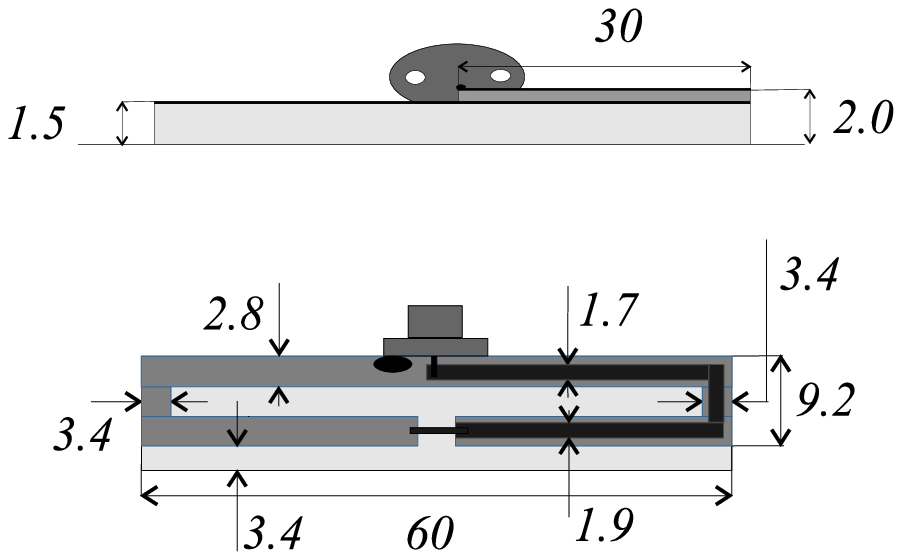}
\end{center}
\caption{A detailed chart of the developed folded-dipole antenna with an integrated symmetrizing device.}
\label{foldeddip_proto_chart}
\end{figure}

The folded dipole and the designed symmetrizing device are inseparable parts
of a microstrip construction. The detailed chart of the construction is given in Figure~\ref{foldeddip_proto_chart}.
The developed radiator is fed as follows. The outer connector of the coax is attached
to the central point of one of the two conductors that form the
folded dipole. The output voltage of the coaxial cable is transmitted
to the feeding point by a microstrip line formed by an additional
strip placed on top of one of the folded dipole conductors. If
the width of the additional strip is smaller than the width of
the dipole strip, and if the separation between the two strips
is much smaller than the width of the folded dipole, the total
construction radiates as a single folded
dipole. That is because the additional strip is effectively
screened by the dipole itself and does not influence
the operation of the antenna. The construction is effectively
symmetric, since the outer connector of the coaxial cable is
attached to a zero-potential point. Moreover, the width of the
additional conductor and the thickness of the insulator can be chosen so that
the additional microstrip line is a 50 Ohm line.
Alternatively, it is possible to use this transmission line segment as
an impedance transformer, if the line characteristic impedance differs from 50 Ohm.

\subsection{Near-field measurements with the prototype antenna}

\begin{figure}[htbp]
\begin{center}
\vspace{2.5cm}
\hspace{-1.5cm}\includegraphics[width=3.5cm]{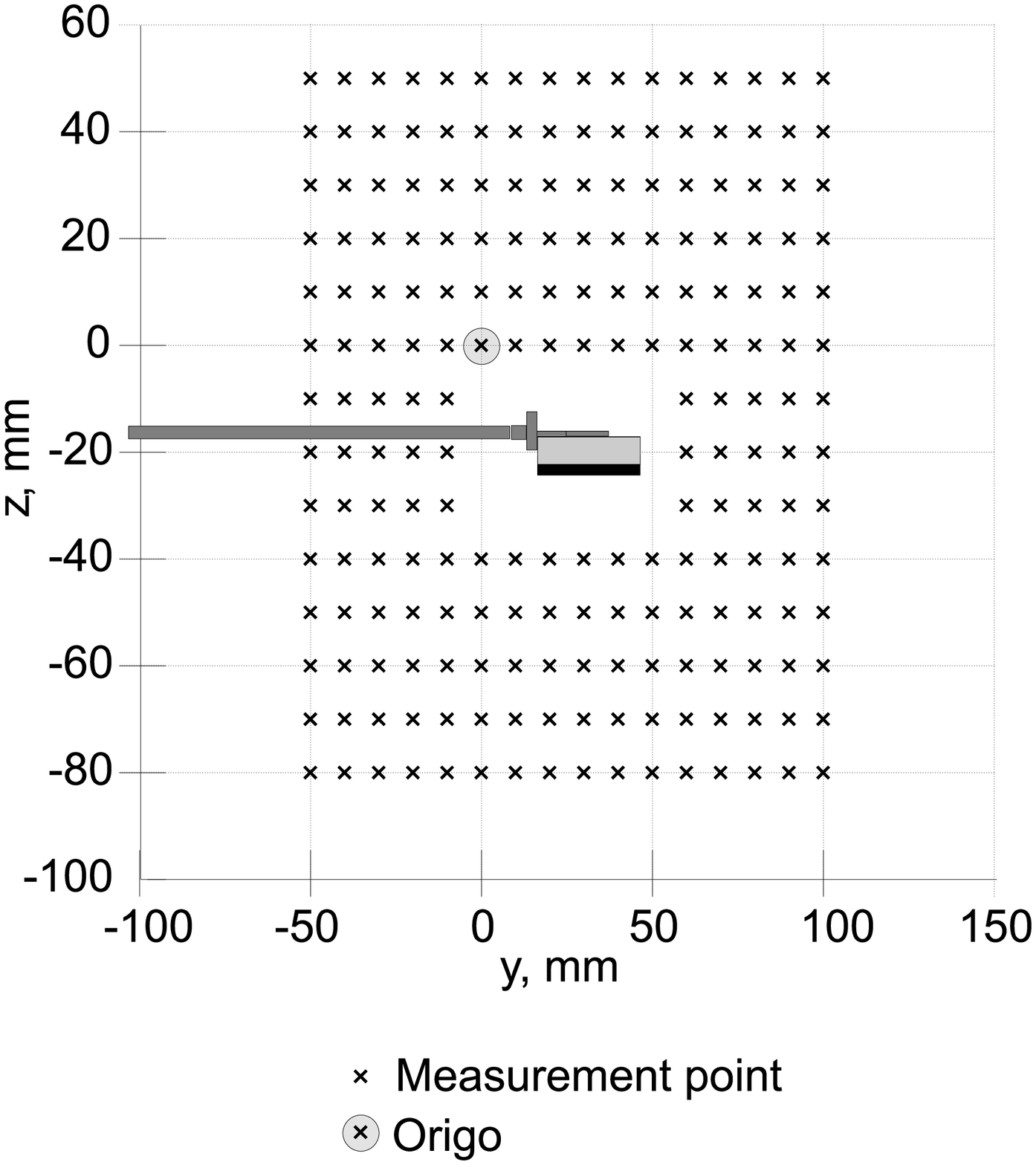}
\end{center}
\caption{The location of the measurement points and the orientation of the prototype antenna under measurement. The feeding coaxial cable is seen on the left.}
\label{measproto_path}
\end{figure}

The prototype antenna consists of a balanced folded dipole over
a mushroom layer of the size $64\times 33$ mm$^2$. The same measurement
procedure as described in Section \ref{half_dip} was used to measure the prototype
antenna. The only difference in the set-up was that it had
no additional metal screen and the cable was directly connected to
the antenna. The location of the measurement points and the orientation of the prototype antenna under measurement are shown in Figure~\ref{measproto_path}.

\begin{figure}[htbp]
\begin{center}
\includegraphics[width=8cm]{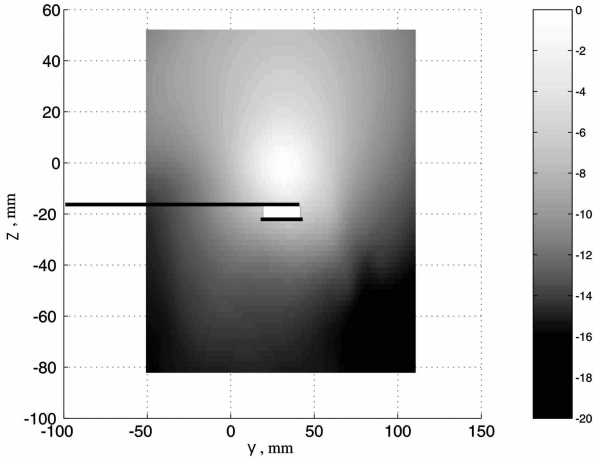}
\end{center}
\caption{Near-field  spatial distribution of the prototype
full-size folded dipole antenna around the radiating system.
The electric field vector relative magnitude, is given in dB.
The position of the antenna and feeding cable are shown. The
radiating structure is located in region $15\ \mbox{mm}<y<45$ mm,
$-25\ \mbox{mm}<z<-17$ mm, see Figure~\ref{measproto_path}.}
\label{fig10}
\end{figure}

\begin{figure}[htbp]
\begin{center}
\includegraphics[width=8cm]{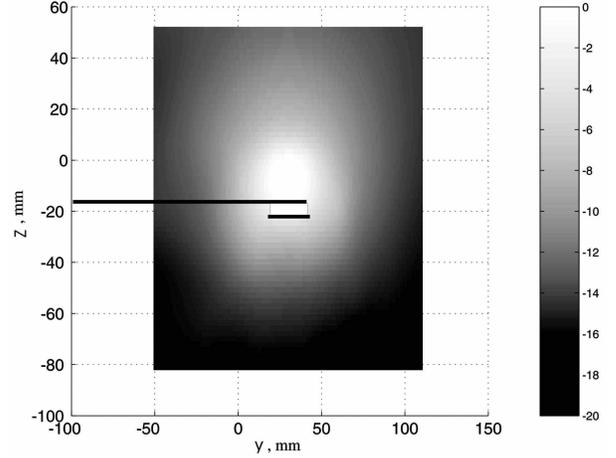}
\end{center}
\caption{The same as in Figure~\ref{fig10} for the magnetic field relative magnitude, dB.}
\label{fig11}
\end{figure}

Test field distributions along the antenna dipole showed that the antenna was well balanced. The operating frequency was 1.77 GHz.
We have measured the field distribution around the antenna and calculated the averaged screening effect taking into account the field at every point shown in Fig. \ref{measproto_path}.
The results depicted in Figures \ref{fig10} and \ref{fig11} are in good agreement with similar ones obtained
in the first experiment with a half of an analogous antenna. A small asymmetry of the electric field distribution is caused by the coaxial cable connected to the dipole. The input return loss at the resonance is $|S_{11}|=-20$ dB, the radiation efficiency  is
73\%, the averaged screening effect $\mbox{ASE}=13$ dB, the input resistance
at the resonance is 61 Ohm, the antenna bandwidth at the level
$-6$ dB (for $S_{11}$) is 9\% (at $-9.5$ dB it is 5.5\%).

We also studied the same radiating element with other
impedance surfaces behind it: Jerusalem-slot screen and a metal plate.
In both these cases we obtained good agreement with the
studies of the half of a similar folded dipole (Section \ref{half_dip}). The final
results are presented in Table~\ref{tab3}.

\begin{table}[h]
\begin{center}
\caption{Main parameters of the three designed prototypes.}
\label{tab3}
\begin{tabular}{|l|l|l|l|}
\hline
                                    & Metal              & Mushrooms & Jerusalem \\
                                    & plate              &           & crosses\\
\hline
Central frequency, GHz              & 1.81               & 1.77      & 1.83\\
\hline
Radiation efficiency$^\dag$, \%            & 76                 & 73        & 88\\
\hline
Total thickness $H$, mm    & 9.35               & 8.3       & 12.3\\
\hline
Bandwidth, \%                       & --                 & 5.5 ($-9.5$ dB) & 4 ($-10$ dB)\\
\hline
Electric field ASE, dB              & 15                 & 13        & 15\\
\hline
Magnetic field ASE, dB              & 15--17             & 12--13    & 13--15\\
\hline
\end{tabular}
\smallskip
\end{center}
$^\dag$ {\footnotesize The efficiency values were measured in the modeling set-up discussed in Section \ref{half_dip}.}
\end{table}

\section{Conclusions}

Compared to the known designs of antennas utilizing high-impedance
surfaces (see, e.g. \cite{2,7}), our prototypes are much superior
in the screening effect. We do not know works in which the
near-field screening effect would be calculated or measured for
antennas operating at 1--3 GHz. In \cite{7} the screening effect
of 7 dB corresponds to the far-zone field, and  in \cite{2} 6 dB
is the front-to-back ratio (also a far-zone parameter). To compare
our results with the basic known ones we should replace ASE
introduced in our paper by its analogue in which the far-zone part
of our measured data is kept. Estimation of the distance to the
far zone is not obvious in this case when the antenna is not
electrically large but cannot be considered as a point source.
Looking at the distance to the far-zone $Z$ for considerably large
antennas \e Z={2G^2\over \lambda}, \f where $G=64$ mm is the
maximal size of our radiating system, and to the same for point
dipoles \e Z={\lambda\over {2\pi}}\f and choosing the larger
value, we estimate that the distance $Z>5$ cm from the antenna
(which corresponds to the $z$-coordinates $z=35\ldots 50$ and
$z=-70\ldots -80$ mm in Fig.~\ref{measproto_path}) can be referred
to the far-zone. Calculating ASE for these distances we obtain 18
dB for the present design against the  known 6-7 dB \cite{2} and
\cite{Pisa}. It is also clearly visible in Figures~\ref{fig6},
\ref{fig7}, \ref{fig10}, \ref{fig11} that the maximal screening
effect in our  measurements corresponds to relatively large
distances. Therefore, we assume that the result of 6 dB for the
front-to-back ratio in the far zone corresponds to a practically
small ASE for the near field and does not witness any considerable
improvement of the near-field pattern of the antenna.

We also have demonstrated increased radiation efficiency (73-88
percent against the known 60 percent). However, the designs of
\cite{2,7} have smaller thicknesses (approximately 3 mm against
our 8-12 mm). Let us explain this point. We have tried to achieve
a compromise between high near-field screening, high efficiency,
and a small thickness with the emphasis on the near-field
screening. Notice, that in other known designs, the resonant
frequency region of the mushroom surface is used. In that regime
the surface operates as a magnetic screen, which means that the
radiating element can be brought very close to the surface without
cancelation of the radiation resistance. This allows a compact
design. But, this leads to a smaller screening efficiency
\cite{1}.  And this point was not studied enough in the known
literature.

Also, the radiation efficiency of an antenna over a
high-impedance surface can be lower than in the regime of
the moderate surface impedance because of stronger fields in the
dielectric substrate. In our design we work at a frequency much
lower than the impedance surface resonance. In this region the
surface impedance is rather low, so that the properties are close
to that of an \textit{electric} screen. For this reason we cannot
bring the primary radiator very close to the surface, and the
thickness is increased, but the screening of the back radiation is
very much improved. The bandwidth is the best for the case of the
mushroom surface. The simple metal plate ($Z_s\approx 0$) can be
used for screening if we neglect the requirement of the broad band
antenna operation (and allow even thicker design).

Summarizing, we can conclude that it has been experimentally
demonstrated that simple artificial impedance surfaces can be
used in the design of antennas for reducing the near
field behind the antenna and increasing the antenna efficiency
without any bandwidth reduction.

\section*{Acknowledgment}
This work was supported in part by Nokia Research Center,
Filtronic LK, and TEKES.

\end{document}